\documentclass[usenatbib,longnamesfirst]{mn2e}
\RequirePackage{graphicx}
\usepackage{hyperref} 

\begin{document}
\title[Taurus molecular cloud]{SCUBA and Spitzer observations of the
Taurus molecular cloud -- pulling the bull's tail}
\author[Nutter, Kirk, Stamatellos \& Ward-Thompson]{D. Nutter\thanks{E-mail: 
David.Nutter@astro.cf.ac.uk}, J. M. Kirk,  D. Stamatellos, D. Ward-Thompson\\
Department of Physics and Astronomy, Cardiff University, Queens Buildings, 
Cardiff, CF24 3AA}

\maketitle

\begin{abstract}
We present continuum data from the Submillimetre Common-User Bolometer Array (SCUBA) on the James Clerk Maxwell Telescope (JCMT), and the Mid-Infrared Photometer for Spitzer (MIPS) on the Spitzer Space Telescope, at submillimetre and infrared wavelengths respectively. We study the Taurus molecular cloud 1 (TMC1), and in particular the region of the Taurus Molecular Ring (TMR). In the continuum data we see no real evidence for a ring, but rather we see one side of it only, appearing as a filament. We name the filament `the bull's tail'. The filament is seen in emission at 850, 450 and 160~$\mu$m, and in absorption at 70~$\mu$m. We compare the data with archive data from the Infra-Red Astronomical Satellite (IRAS) at 12, 25, 60, 100~$\mu$m, in which the filament is also seen in absorption. We find that the emission from the filament consists of two components: a narrow, cold ($\sim$8~K), central core; and a broader, slightly warmer ($\sim$12~K), shoulder of emission. We use a radiative transfer code to model the filament's appearance, either in emission or absorption, simultaneously at each of the different wavelengths. Our best fit model uses a Plummer-like density profile and a homogeneous interstellar dust grain population. Unlike previous work on a similar, but different filament in Taurus, we require no grain coagulation to explain our data.
\end{abstract}


\section{Introduction}

The Taurus molecular cloud is one of the nearest star forming regions, at a 
distance of $\sim$140pc \citep{1978ApJ...224..857E}. As such it has often 
been the target of observational studies into the processes of molecular 
cloud evolution and star formation. The large scale structure of the 
molecular cloud has been mapped using the emission from CO 
\citep{1987ApJS...63..645U,1996ApJ...465..815O,2005prpl.conf.8268G}, 
and also the extinction of starlight 
\citep{1999A&A...345..965C,2002ApJ...580L..57P,2005PASJ...57S...1D}.

The cloud has been mapped to study the formation mechanism of molecular 
clouds \citep{1999ApJ...527..285B}, and the nature of the turbulence that 
pervades them \citep[see review by][]{2004ARA&A..42..211E}. On smaller 
scales, detailed studies of the densest regions of the cloud have greatly 
increased our knowledge of the star-formation process itself 
\citep{2000ApJ...537L.135W,2001A&A...365..440M,2002ApJ...575..950O,
2004A&A...427..651D,2005MNRAS.360.1506K}.

Taurus differs from other nearby sites of star formation, in that it is 
forming stars in a quiescent manner. There are no newly formed, massive 
stars injecting energy into the cloud; and compared to more dynamic 
star-formation regions like Orion or $\rho$-Ophiuchus, each of the 
star-forming cores are widely separated 
\citep[e.g.][]{1998A&A...336..150M,1999MNRAS.305..143W}. The cloud also 
appears to have a core mass function which differs from that found in other 
regions \citep{2002ApJ...575..950O,2004A&A...419..543G}.

\begin{figure*}
\includegraphics[angle=0,width=175mm]{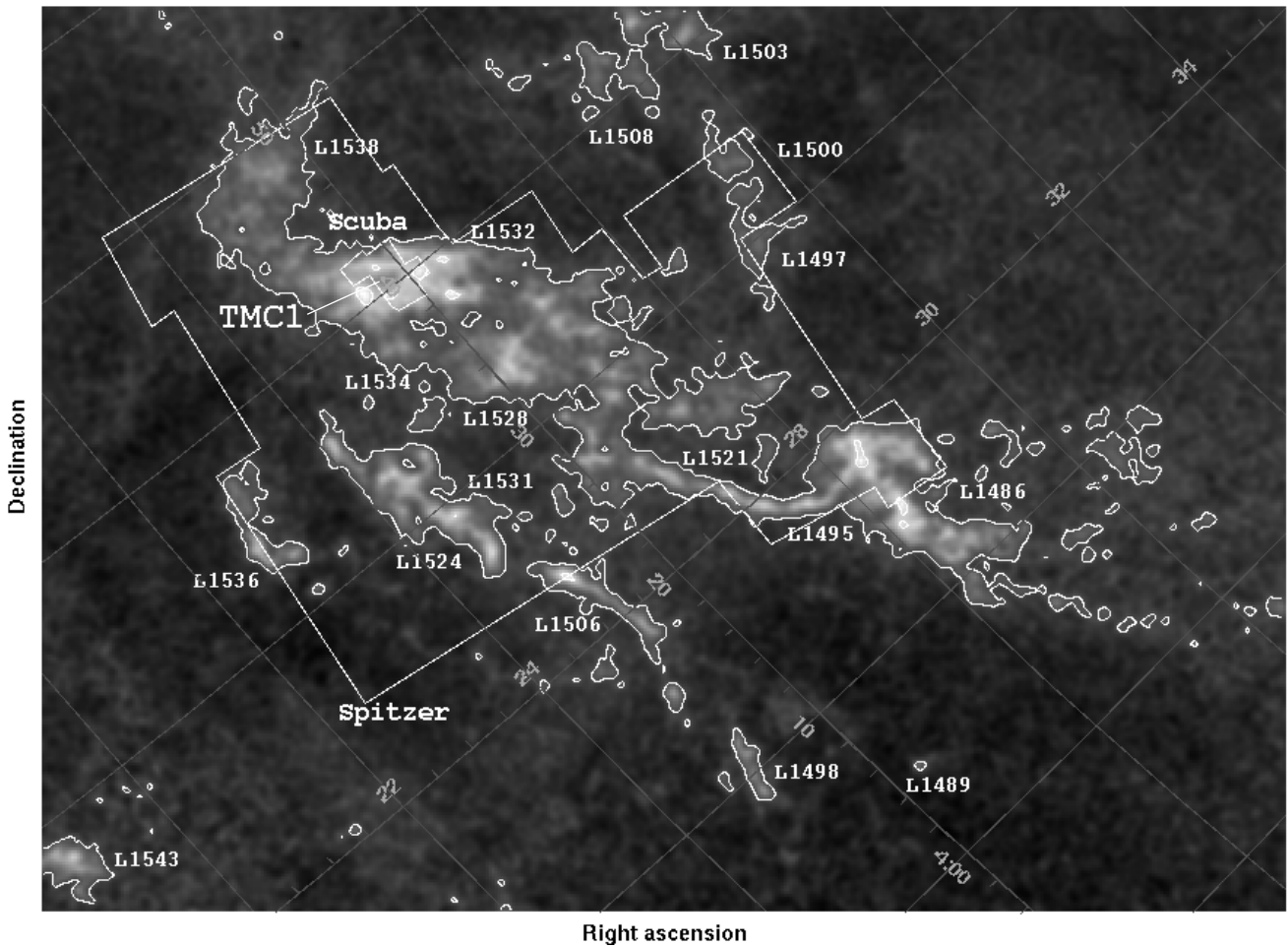}
\caption{The Taurus molecular cloud seen in extinction 
\citep{2005PASJ...57S...1D}. The Av=1 and Av=5 contours are shown. 
TMC1 is marked, as are a number of Lynds dark clouds 
\citep{1962ApJS....7....1L}. The outlines of the regions mapped by
MIPS at 160$\mu$m (large box), 
and SCUBA at 850$\mu$m (small box), are shown. The SCUBA data 
are centred around TMC1.}
\label{taurus_overview}
\end{figure*}

\begin{figure*}
\begin{minipage}[t]{87mm}
\vspace{0pt}
\includegraphics[angle=0,width=80mm]{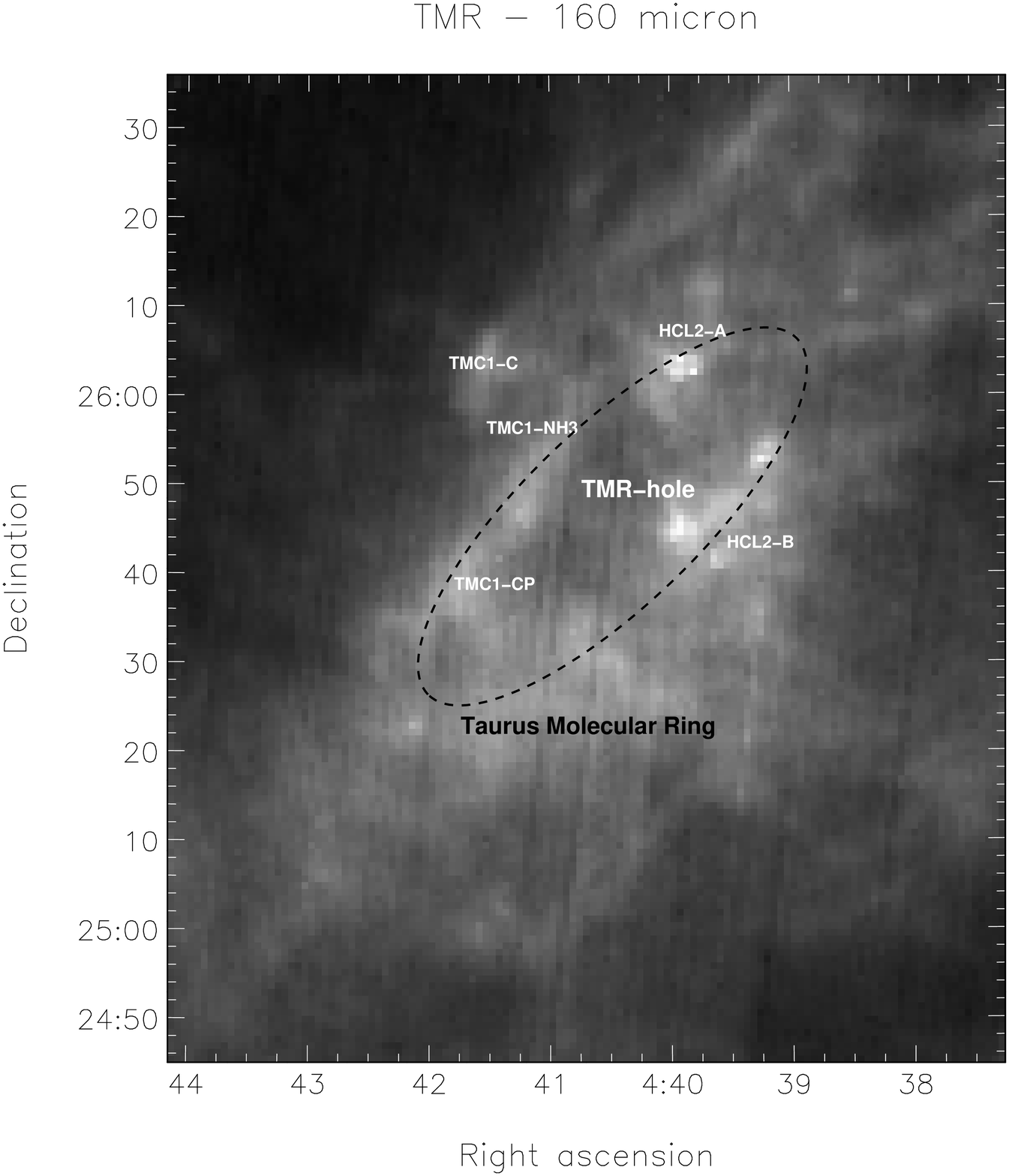} 
\end{minipage}
\begin{minipage}[t]{87mm}
\vspace{0pt}
\includegraphics[angle=0,width=80mm]{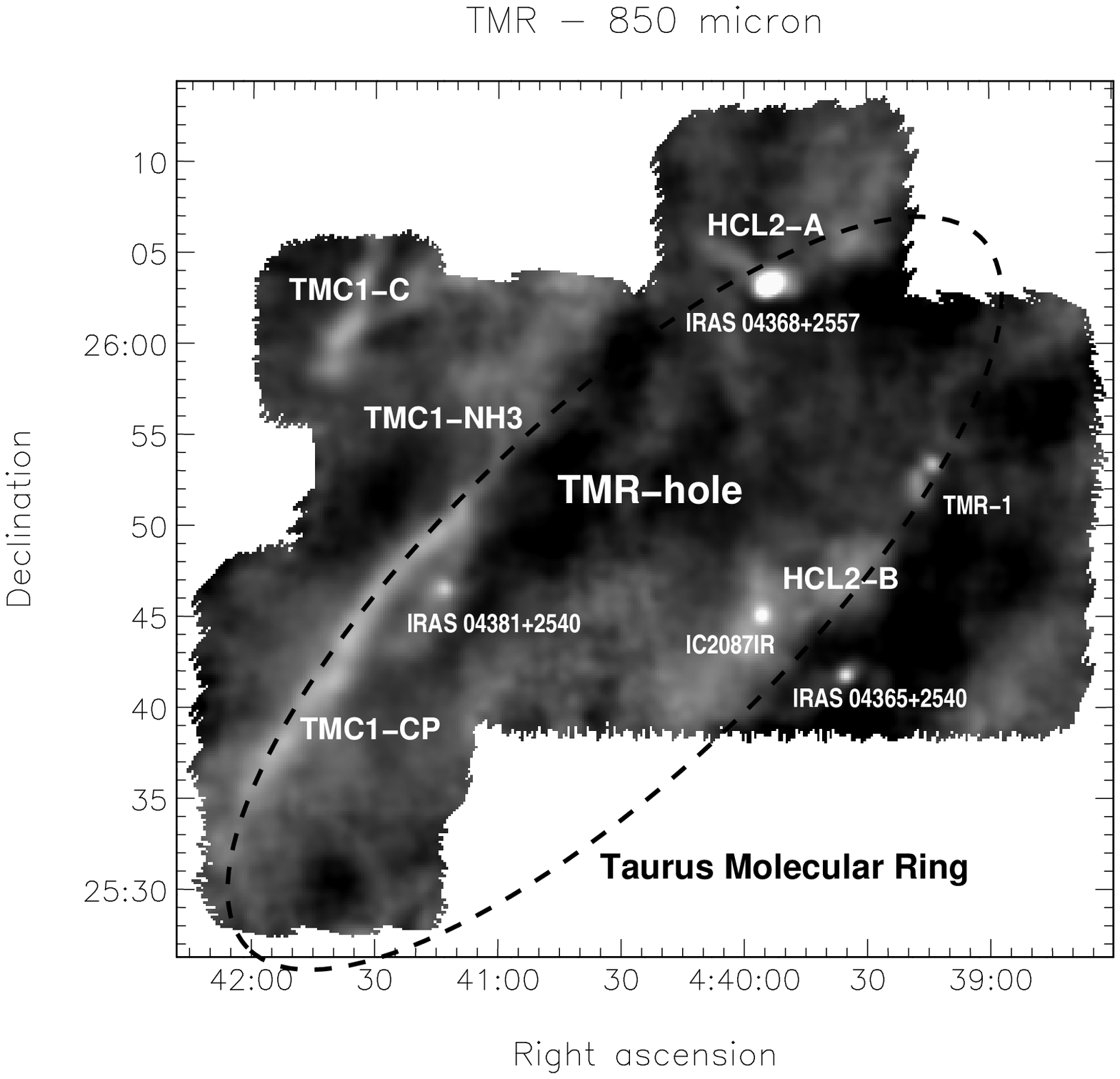}
\end{minipage}
\caption{A close-up of the TMC1 region at (a) 160~$\mu$m, and (b) 850~$\mu$m. 
The bright components of TMC1 and HCL2 are marked
(TMC1-CP and TMC1-NH3 refer to the 
cyclopolyyne and ammonia peaks respectively -- \citealp{1988A&A...196..194O}).
Some bright IRAS sources are also marked. The 
approximate locus of the TMR is shown as a dashed line.
The ring structure is not very clear in the continuum images.
Instead we see a filament coincident with the eastern portion of the ring,
and simply a collection of point sources and clumps in the western half.
We here name the filament `the bull's tail'.} 
\label{taurus_overview_zoom}
\end{figure*}

\begin{figure*}
\begin{minipage}[t]{87mm}
\includegraphics[angle=0,width=87mm,height=94mm]{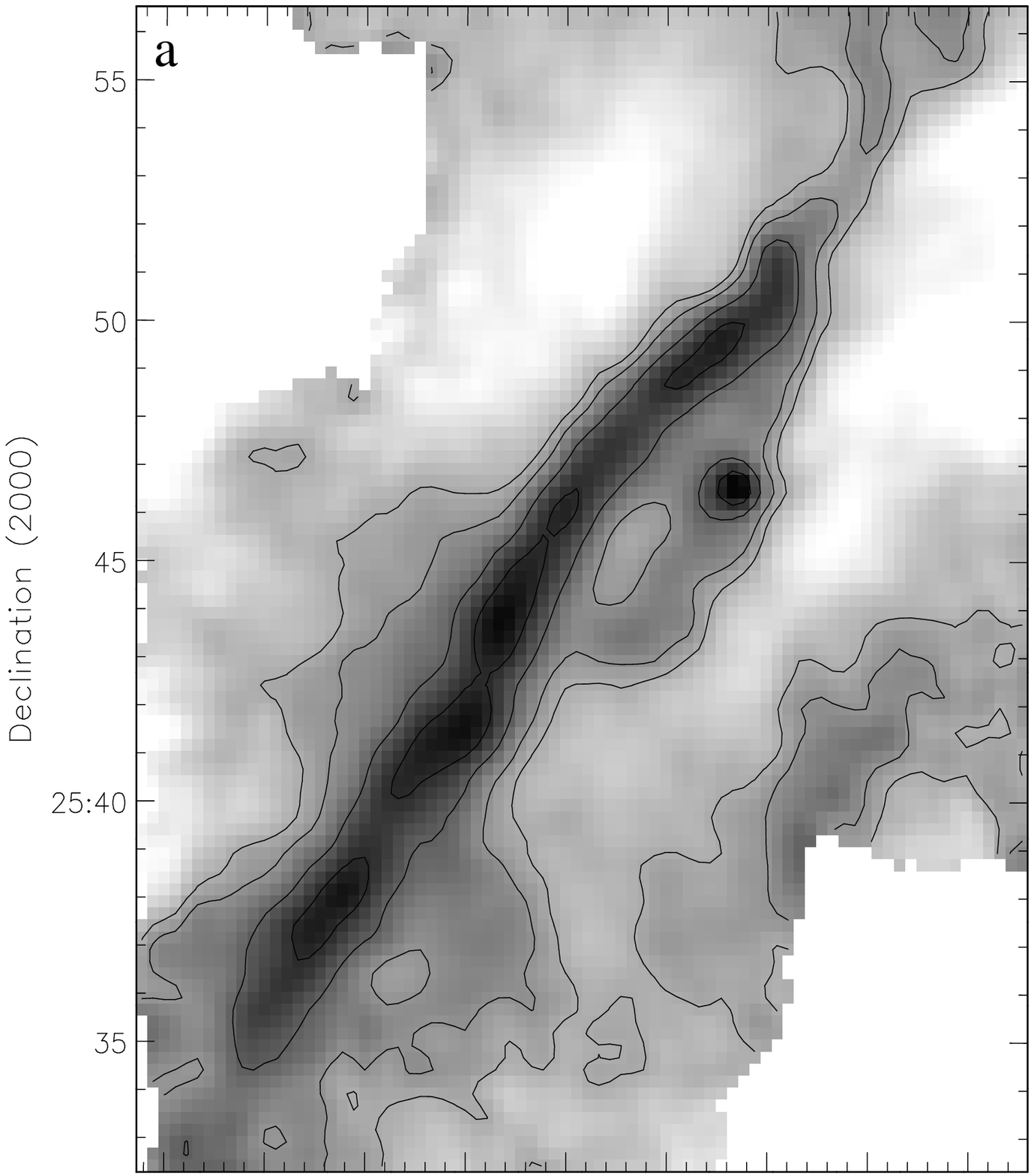} 
\end{minipage}
\begin{minipage}[t]{87mm}
\includegraphics[angle=0,width=75.8mm,height=94mm]{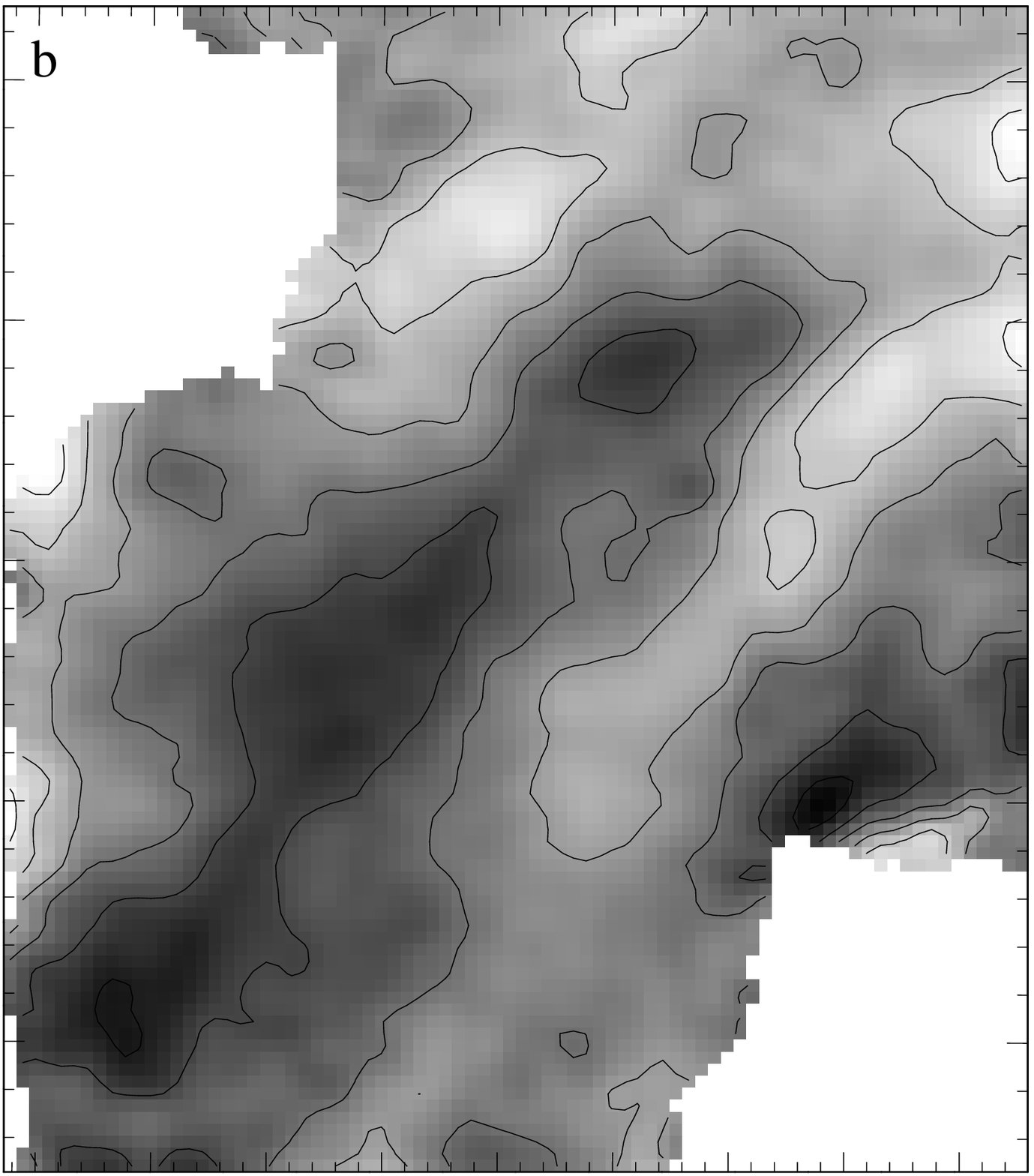}
\end{minipage}
\begin{minipage}[t]{87mm}
\vspace{0pt}
\includegraphics[angle=0,width=87mm,height=101mm]{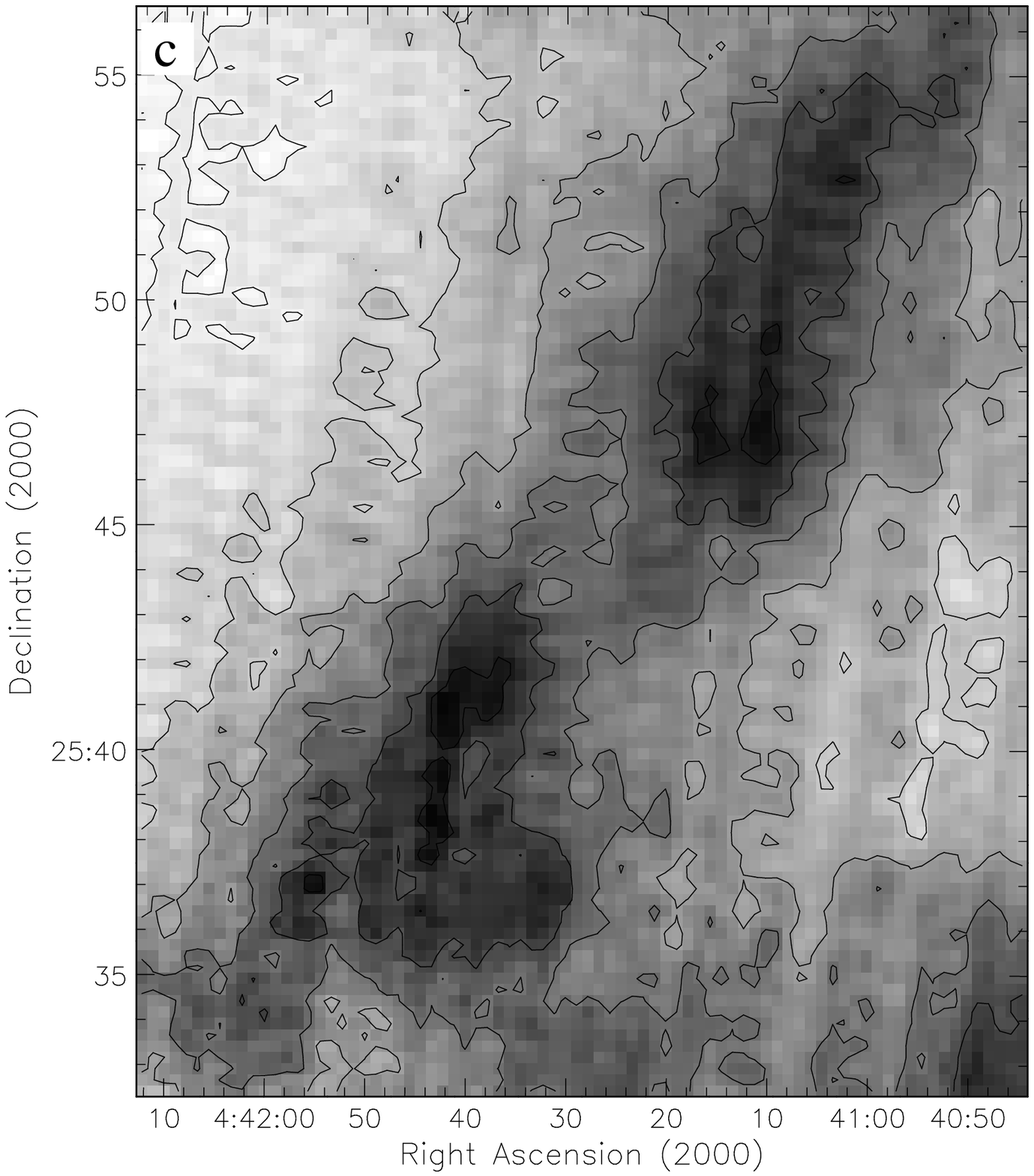}
\end{minipage}
\begin{minipage}[t]{87mm}
\vspace{0pt}
\includegraphics[angle=0,width=75.8mm,height=101mm]{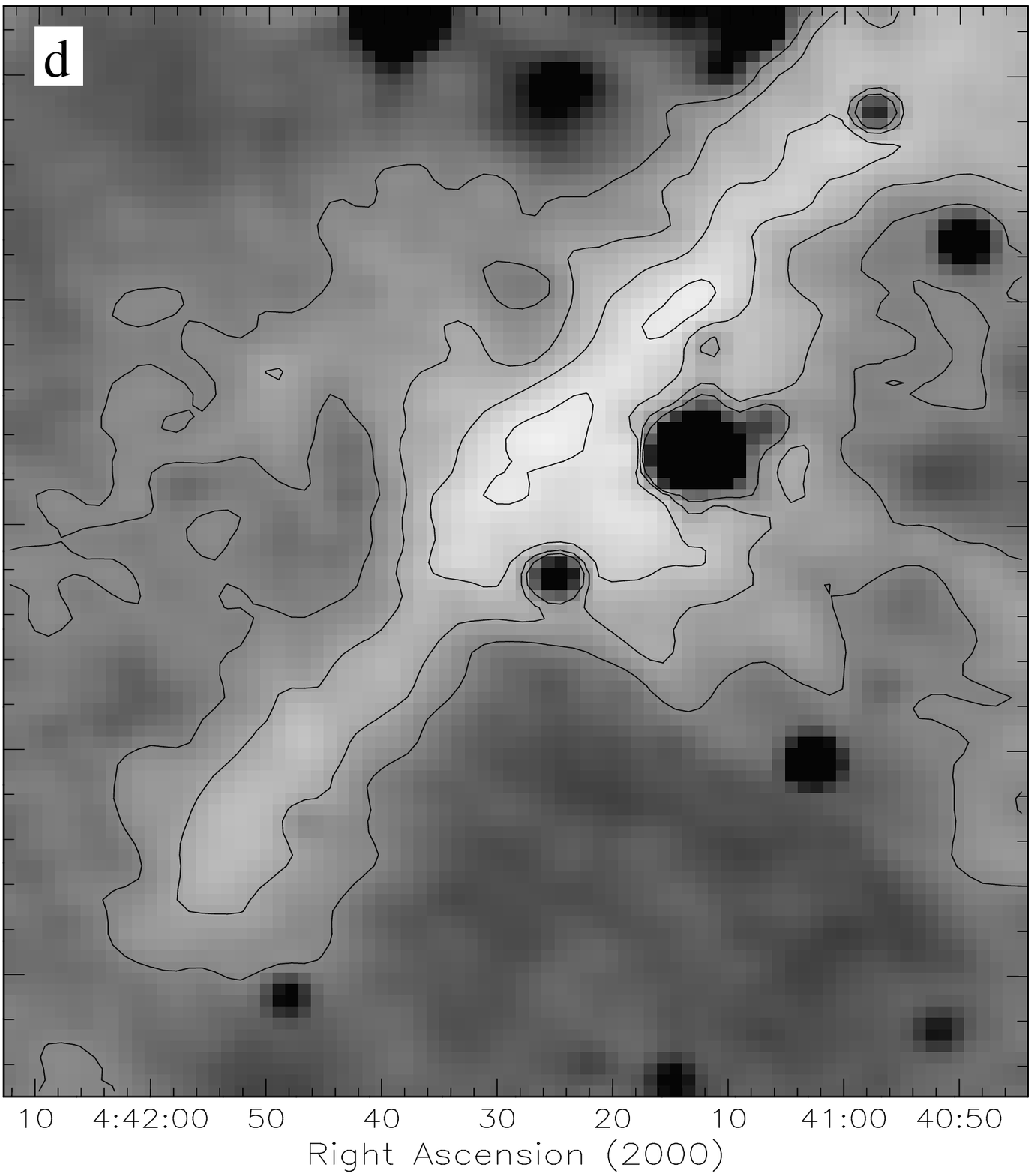}
\end{minipage}
\caption{An even closer view of the bull's tail filament 
at (a) 850~$\mu$m, (b) 450~$\mu$m, 
(c) 160~$\mu$m, and (d) 70~$\mu$m. The filament is seen in emission at the 
three long wavelengths, and in absorption at 70~$\mu$m. Each of the maps 
has been smoothed to match the 45-arcsec angular resolution of the MIPS 
160-$\mu$m data. The remaining differences in the width of the filament 
are therefore intrinsic to the source itself. The filament is narrowest 
at 850 and 70~$\mu$m, where there is good agreement between the emission
at 850 and absorption at 70~$\mu$m. The filament appears broader at the
other two wavelengths.} 
\label{taurus_maps1}
\end{figure*}

\begin{figure*}
\includegraphics[angle=270,width=175mm]{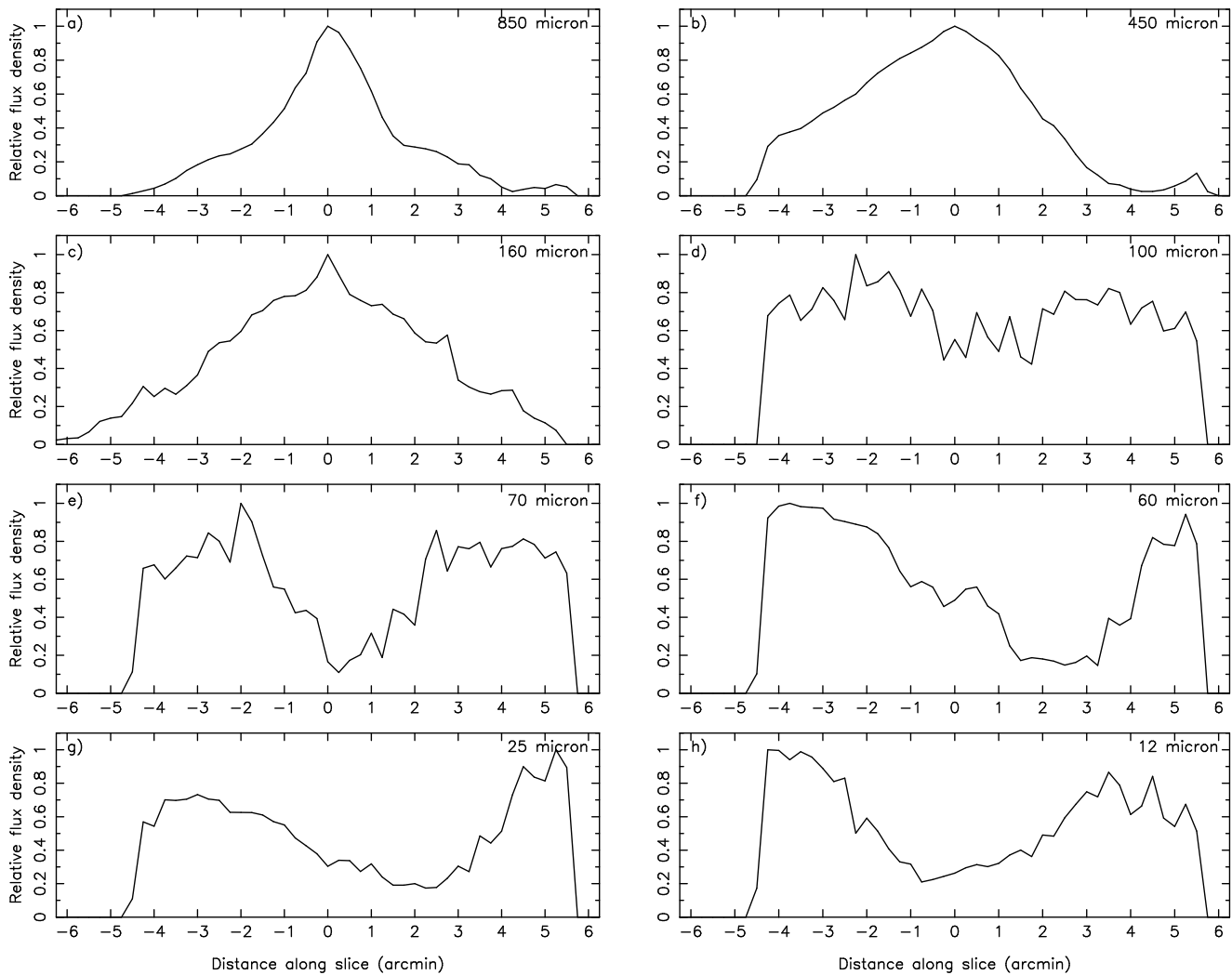}
\caption{One-dimensional cuts through the data perpendicular to the
filament at each of the eight wavelengths studied. 
Each cut has been normalised to its own peak.}
\label{slices}
\end{figure*}

The far-infrared and submillimetre continuum represent an efficient means of
detecting cold star-forming cores in molecular clouds, even in energetic
regions such as Orion \citep{1995MNRAS.274.1219W,1996A&A...314..477B}.
Figure~\ref{taurus_overview} shows the Taurus molecular cloud as seen in 
extinction \citep{2005PASJ...57S...1D}, with the contours showing the visual 
extinctions of Av=1 and Av=5. A number of the Lynds clouds which are 
associated with the Taurus molecular cloud are labelled.

The majority of 
the cloud has been mapped in the far-infrared by Spitzer \citep[][for more 
details, see Section \ref{spitzer}]{2007prpl.conf..329G}. The region
mapped by Spitzer is 
outlined in white on Figure~\ref{taurus_overview}. A
somewhat smaller region, centred on Taurus molecular cloud 1 (TMC1),
has been mapped by SCUBA on the JCMT (see 
Section \ref{scuba}). This smaller region is also outlined in white on 
Figure~\ref{taurus_overview}.
The densest part of the cloud is TMC1. Together with Heiles Cloud~2
\citep[HCL2;][]{1968ApJ...151..919H} this makes up the Taurus Molecular Ring 
(TMR) \citep[see][]{2004A&A...420..533T}. 

Figures~\ref{taurus_overview_zoom}(a) and (b) show the region around TMC1 at 
160 and 850~$\mu$m respectively, with the brighter components of Heiles 
Cloud 2 and TMC1 marked. In addition, the approximate locus of the TMR 
is also marked. For more details about these data, see 
Section~\ref{observations}. However, we note here that the molecular ring
is not really seen in the continuum. Instead, we see a filament in the eastern
half and a series of point-like and extended sources scattered roughly
around the western half. We therefore challenge the existence of the ring
as a single coherent entity, as that is not what is seen in these data.
We here name the filament `the bull's tail'.

In this paper, we present a study of a portion of TMC1 using infrared and 
submillimetre data between 12 and 850~$\mu$m. This range of wavelengths 
straddles the peak of the spectral energy distribution (SED) for the cold 
dust that is found in the densest regions of molecular clouds. Infrared and 
sub-mm data are therefore a powerful tool for breaking the degeneracy of 
temperature and density that is present when only sub-mm data are considered. 

\section{Observations}\label{observations}
\subsection{SCUBA Data}\label{scuba}

The submillimetre data presented in this study were obtained using the 
Submillimetre Common User Bolometer Array (SCUBA
-- \citealp{1999MNRAS.303..659H}) on the James Clerk 
Maxwell Telescope (JCMT). SCUBA takes observations at 450 and 
850~$\mu$m simultaneously through the use of a dichroic beam-splitter. The 
telescope has a resolution of 8 arcsec at 450$~\mu$m and 14 arcsec at 
850~$\mu$m. The data presented here were acquired from the JCMT data 
archive, operated by the Canadian Astronomy Data Centre. A sub-set of 
these data has been published previously \citep{2007ApJ...657..838S}, 
though all of the data have been re-reduced by us using a consistent 
method for the purposes of this study.

The observations were carried out over 12 separate nights between 
December 1998 and January 2005 using the scan-map observing mode. 
A scan-map is made by scanning the array across the sky, using a 
scan direction of 15.5$^\circ$ from the axis of the array in order 
to achieve Nyquist sampling. The array is rastered across the sky to 
build up a map several arcminutes in extent.

Time-dependent variations in the sky emission were removed by chopping 
the secondary mirror at 7.8 Hz. The size of a scan-map is larger than 
the chop throw, therefore each source in the map appears as a positive 
and a negative feature. In order to remove this dual-beam function, each 
region is mapped six times, using chop throws of 30, 44 and 68 arcsec in 
both RA and Dec \citep{1995mfsr.conf..309E}. The dual-beam function is 
removed from each map in Fourier space by dividing each map by the Fourier 
transform of the dual-beam function, which is a sinusoid. The multiple 
chop-throws allow for cleaner removal of the dual beam function in Fourier 
space. The maps are then combined, weighting each map to minimise the noise 
introduced at the spatial frequencies that correspond to zeroes in the 
sinusoids. Finally the map is inverse Fourier transformed, at which point 
it no longer contains the negative sources \citep{SURF}.

The submillimetre zenith opacity at 450 and 850$~\mu$m was determined using 
the `skydip' method and by comparison with polynomial fits to the 1.3~mm sky 
opacity data, measured at the Caltech Submillimeter Observatory 
\citep{2002MNRAS.336....1A}. The sky opacity at 850$~\mu$m varied from 
0.18 to 0.59, with a median value of 0.27. These correspond to a 450$~\mu$m 
opacity range of 0.81 to 3.5, and a median value of 1.4. 

The data were reduced in the normal way using the SCUBA User Reduction 
Facility, SURF \citep{SURF}. Noisy bolometers were removed by eye, and the 
baselines, caused by chopping onto sky with a different level of emission, 
were removed using the {\em \small MEDIAN} filter. Calibration was performed 
using observations of the planets Uranus and Mars, and the secondary 
calibrator CRL618 \citep{1994MNRAS.271...75S} taken during each shift. 
We estimate that the absolute calibration uncertainty is $\pm 5\%$ at 
850$~\mu$m and $\pm 15\%$ at 450~$\mu$m, based on the consistency and 
reproducibility of the calibration from map to map. 
The resolution of the SCUBA data was degraded using a Gaussian smoothing 
routine to a 45 arcsec beam, to allow a direct comparison of these data
with the Spitzer data.

\subsection{MIPS Data}\label{spitzer}

The Multiband Imaging Photometer for Spitzer 
\citep[MIPS;][]{2004ApJS..154...25R} is the far-IR camera on board the 
Spitzer Space Telescope \citep{2004ApJS..154....1W}. It operates at 24, 
70 and 160~$\mu$m, and has a diffraction-limited resolution of 6, 18 and 
40~arcsec respectively at the three wavelengths. 

Spitzer legacy surveys have been used to map most of the star-forming regions within 0.5~kpc and  
hence produce a complete mid-infrared record of the nearby large molecular clouds.
These legacy surveys include `Cores to Disks' \citep[c2d --][]{2003PASP..115..965E}, 
the `Spitzer Gould Belt Legacy Survey' \citep{GB_pasp}, and the `Taurus-2 Spitzer Legacy 
Project'\footnote{\url{http://spider.ipac.caltech.edu/staff/dlp/taurus/}} 
\citep{2007prpl.conf..329G}. The Spitzer archive data used here were observed as 
part of the `Taurus-2 Spitzer Legacy Project'.

The {\small LEOPARD} software tool \citep{LEOPARD} was used to search the 
Spitzer data archive for MIPS scan-map data coincident with the 
position of the SCUBA data. 160-$\mu$m unfiltered basic calibrated datasets 
1122-6112/6368/6624/6880 from pipeline version S14.4 were identified and 
downloaded. The data were filtered to remove the frames taken directly after
stimulator-flashes (calibration events that can leave bright residuals). The 
data were re-gridded with the MOPEX software package into an image with 
15-arcsec-diameter pixels \citep{2005ASPC..347...81M}. This image was then 
iteratively cleaned and filled to reject spurious pixel artefacts as 
described in \citet{2007MNRAS.375..843K}.

Rather than process the 70-$\mu$m data ourselves we have made use of data 
released by this Legacy Project. These data come from the same original 
dataset, but have been processed to an enhanced level by the Legacy 
Project team. The data were originally reduced with pipeline version S11.1,
and finally smoothed to the same 40-arcsec resolution as the 160-$\mu$m data.

The chopping that is employed by the JCMT to remove the effects of the 
atmosphere also has the effect of limiting the sensitivity to spatial 
scales larger than a few times the largest chop throw. Spitzer does not 
need to chop, and so does not suffer from this. In order to directly 
compare the MIPS and SCUBA data, synthetic chops with the same properties 
as the SCUBA chops were added to the MIPS data using the {\em \small SURF} 
routine {\em \small ADD\_DBM}. These chops were then removed using the same 
method as that employed for the SCUBA data. The SCUBA and MIPS data therefore 
have the same angular resolution, and are sensitive to structures on the same 
spatial scales and can thus be compared directly.

\subsection{IRAS Data}\label{iras}

The Infrared Astronomical Satellite \citep[IRAS;][]{1994STIN...9522539W} was 
a 60-cm space-borne telescope which carried out all-sky surveys at 12, 25, 
60 and 100~$\mu$m. IRAS had an angular resolution of 20, 20, 60 and 120 
arcsec respectively at the four wavelengths. ISSA data for TMC1 were retrieved 
from the IRAS archive
using the {\em skyview} interface \citep{1998IAUS..179..465M}. 
In order to remove large scale gradients which were present in the IRAS data, 
they were smoothed using a 5-arcmin Gaussian function. This 
smoothed map was subsequently subtracted from the data, thus removing the 
large-scale structure, and also making the data directly comparable with the
chopped data.

\section{Results}

Figures~\ref{taurus_maps1}(a)--(d) show the bull's tail at 850, 450, 160 and 
70~$\mu$m respectively. The three long wavelength maps show the 
filament in emission, whereas the 70~$\mu$m map shows the filament in 
absorption against the warmer background of the surrounding cloud.
This is therefore clearly a dense structure, for it to be seen in absorption
at a wavelength as long as 70~$\mu$m.

Figure~\ref{taurus_maps1} shows that the filament is much narrower at 
850 than at 450 or 160~$\mu$m. In addition, the absorption at 70~$\mu$m
is also very narrow, and matches quite well to the 850-$\mu$m emission.
We interpret this as the 
filament having a cold dense inner region, where the extinction is the 
highest, surrounded by a warmer, less dense outer `jacket'. This is 
explored further in Section~\ref{empirical_model}.

Figure~\ref{slices} shows a series of one-dimensional cuts through the
data of Figure~\ref{taurus_maps1}, in a direction perpendicular to the
long axis of the filament. In addition, Figure~\ref{slices} shows
the equivalent cuts through the IRAS data at 12, 25, 60 \& 100~$\mu$m.
Figure~\ref{slices} shows that 
there is a transition at approximately 100 $\mu$m, longward of which the 
filament is seen in emission. At wavelengths shorter than 100~$\mu$m 
the filament is seen in absorption. At 100~$\mu$m there is a hint that
the filament is seen in absorption, although the data are quite noisy.
Figures~\ref{slices}(a)--(c) also confirm that the filament is narrower 
at 850~$\mu$m than at 450 or 160~$\mu$m. We investigate 
this effect in Section \ref{empirical_model}.

\section{Analysis}

\subsection{Empirical Model}\label{empirical_model}

\begin{figure}
\includegraphics[angle=270,width=87mm]{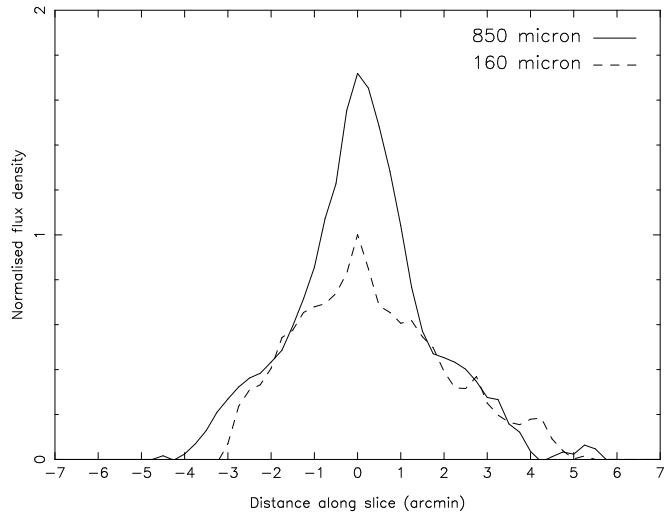}
\caption{The 850- and 160-$\mu$m cuts normalised to a 12-K blackbody 
(see text for details). Note that the two curves match up well in the
`shoulders' of the two profiles, but that the 850-$\mu$m data have an
additional narrow component of extra emission on the peak of the filament.}
\label{slice_160_850}
\end{figure}

Figure~\ref{slice_160_850} shows the 850- and the 160-$\mu$m profiles
overlaid. The two profiles have been normalised to a blackbody with a 
temperature of 12~K. To do this, the observed 850- and 160-$\mu$m profiles were divided
by the flux-density which would be emitted by a 12~K blackbody with a 
160-$\mu$m flux-density equal to the observed value at the filament centre. 
The two profiles would have the same peak value of 1 
if the data were due to a single emitting source at 12~K. We use a blackbody function
in this instance, as it has fewer free parameters than the modified blackbody used 
in Section~\ref{temp_analysis}.

In the outer parts of the filament, there is quite a good agreement between the 
two profiles, indicating that the same dust is emitting at both wavelengths,
and that the 850-$\mu$m data also match this model quite well in the
`shoulders' of the profile. However, there is an excess of emission at 850~$\mu$m in the centre of the
profile. This shows that there is a second dust 
component in the centre of the filament which is colder than 12~K, which
therefore emits more strongly 
at 850~$\mu$m. In Section \ref{model} we discuss specific models of 
the filament, but first we attempt a simple spectral energy
distribution (SED) analysis of the two components.

\subsection{Temperature Analysis}\label{temp_analysis}

We measured the total flux density from the filament at each of the three
wavelengths in which it is seen in emission. We also separated out the
flux density of the central component from that of the outer shoulder, 
or jacket, by fitting two components to the various profiles, such as those
seen in Figure~\ref{slice_160_850}.

\begin{figure}
\begin{minipage}{87mm}
\vspace{0pt}
\includegraphics[angle=0,width=80mm]{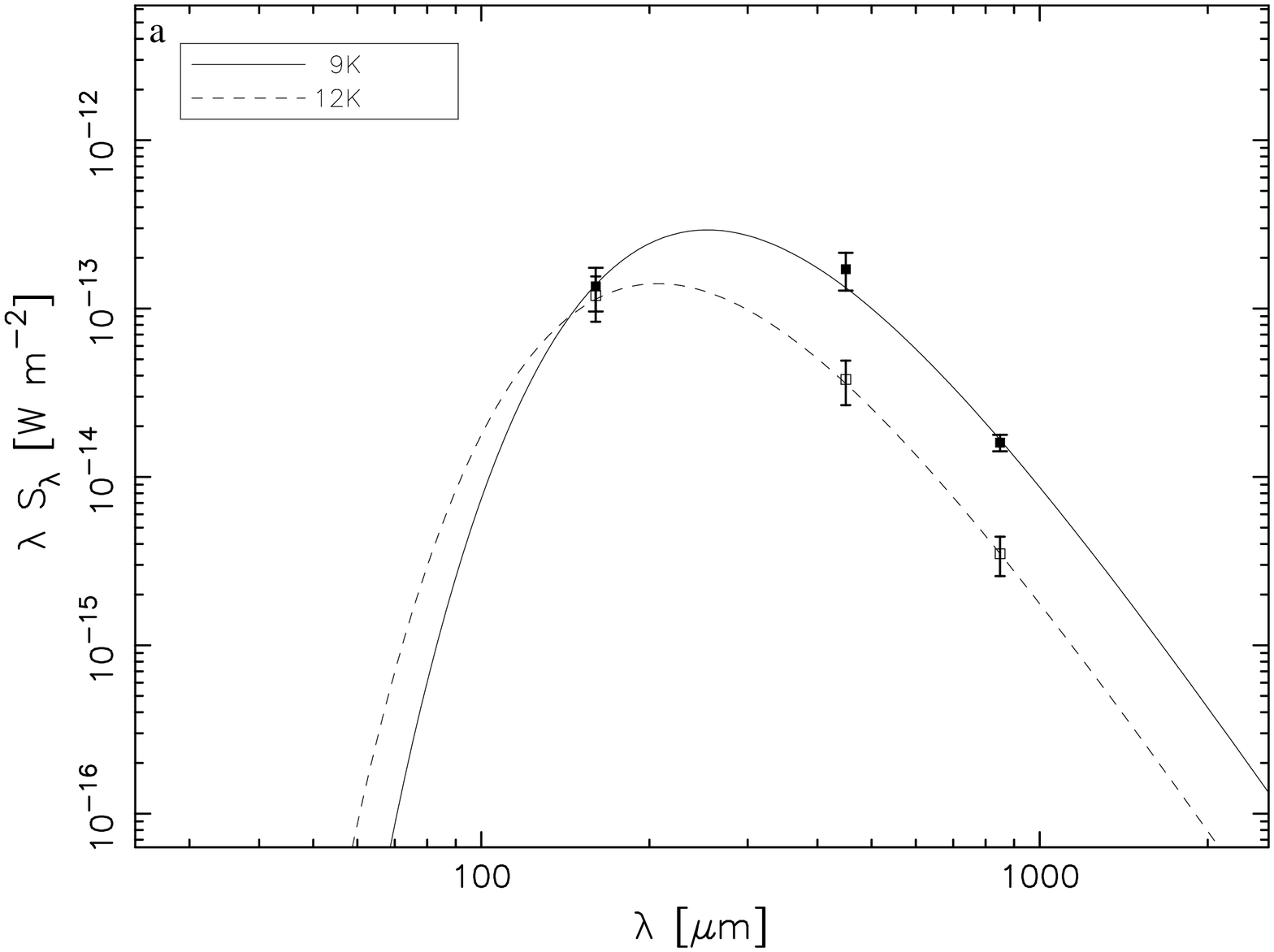} 
\end{minipage}
\begin{minipage}{87mm}
\vspace{0pt}
\includegraphics[angle=0,width=80mm]{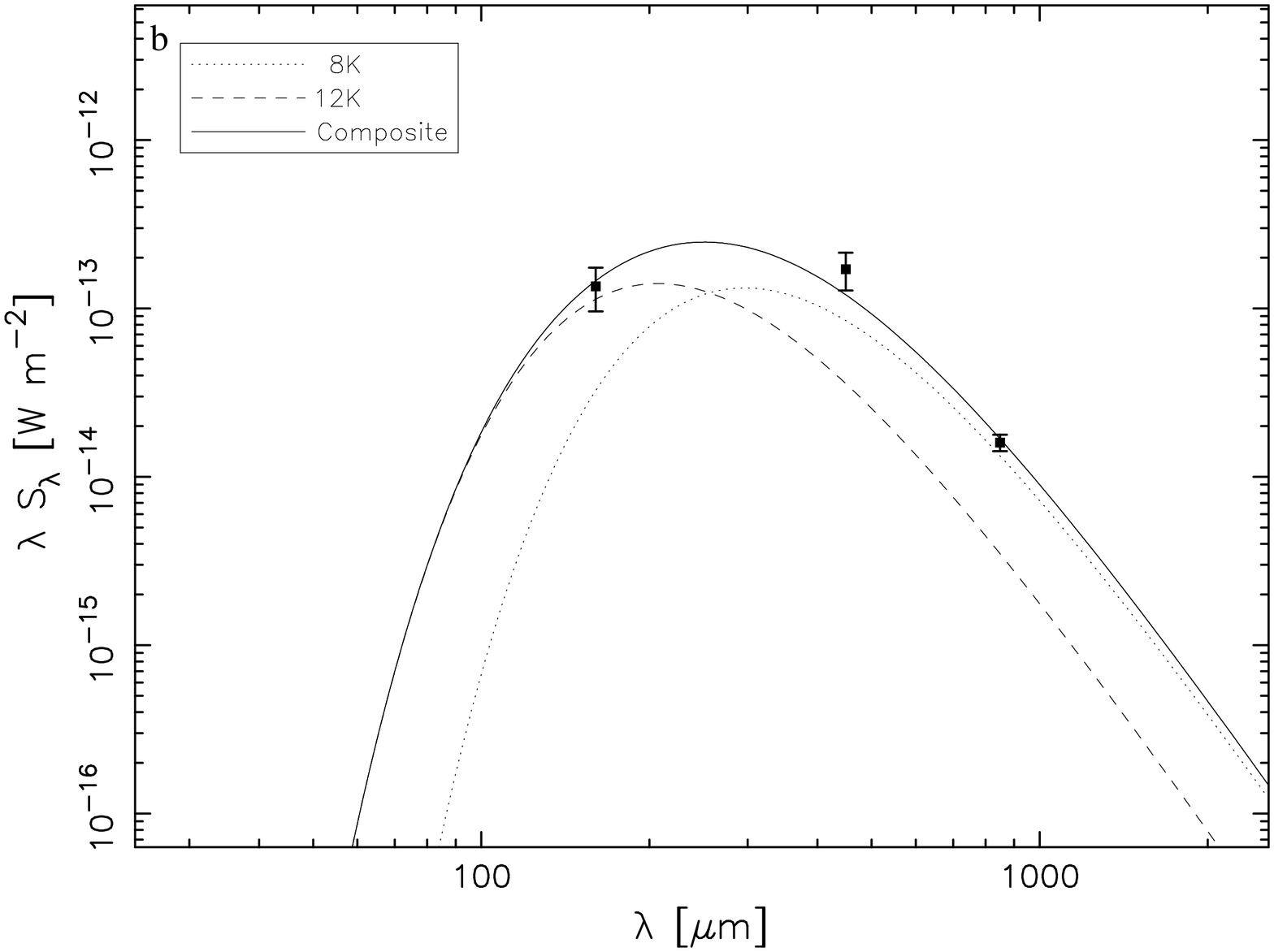}
\end{minipage}
\caption{(a) Flux densities of the central component of the filament (filled 
squares) and the outer component (open squares) at each of the three
wavelengths at which the filament is seen in emission. Spectral energy 
distribution fits in the form of modified blackbody curves are also shown 
for both the central filament (solid line -- 9~K) and the outer filament 
(dashed line -- 12~K). (b) Flux densities for the central component of the
filament only. A two-component SED for the central part of the filament is
also shown. The dashed line represents the same 12-K fit as before, but now
the dotted line shows a fit to the residual flux densities at the two longer 
wavelengths, which corresponds to an 8-K modified blackbody.
The solid line shows the sum of the two.} 
\label{seds}
\end{figure}

Figure~\ref{seds}(a) shows the flux densities of the
filament centre (solid squares) and the outer shoulder
(hollow squares) together with two modified blackbody fits to the data
in the normal manner \citep[see, e.g.][]{2002MNRAS.329..257W}.
Following \citet{2007MNRAS.375..843K}, the 
parameters $\beta$ and the critical wavelength ($\lambda_c$) are constrained 
to be equal to 2 and 50~$\mu$m respectively. The best fits
to the data are found to be 12~K for the 
outer filament, and 9~K for the filament centre.

We assume when we look along a line of sight
towards the centre of the filament, that as well as the cold central
filament, we also see the outer component (in front of, and 
behind the central filament). Therefore, to model the emission from the 
central component, we must also take account of the outer component. Hence
we must fit a two-temperature SED to the emission from the central component.

Figure~\ref{seds}(b) shows the flux densities of the central component
of the filament fitted by two modified blackbodies. We constrain the warmer
component to be 12~K, as discovered in Figure~\ref{seds}(a), and fit the
residual emission with a second component. In this way we find 
that the best fit temperature for the cold dust is actually 8~K. The resulting two-part SED is plotted on Figure~\ref{seds}(b).
Hence we see that the best fit to the data from a simple two-component SED
analysis is for a central cold filament at 8~K, surrounded by a broader
and warmer component at 12~K.

\subsection{Comparison with previous work}

\citet{2004A&A...420..533T} used the
IRAS 60- and 100-$\mu$m data, together with  200-$\mu$m 
data from the Infrared Space Observatory (ISO)
to determine the temperature of TMC1-CP, which is 
roughly coincident with 
the filament that we are studying (see 
Figure~\ref{taurus_overview_zoom}). 

They estimated a temperature of 11.4~K, which 
is consistent with the 12~K that
we have measured in the outer part of the filament.
\citet{2004A&A...420..533T} also found a similar temperature from studies of ${\rm NH_3(1,1)}$ and ${\rm (2,2)}$. \citet{2003A&A...398..551S} used the balloon-borne 
PRONAOS telescope to observe the L1506 cloud in 
Taurus (see Figure~\ref{taurus_overview}). 
They measured a temperature of 12.1~K. 

\citet{2007ApJ...657..838S} observed the TMC1-C 
region using SCUBA data at 450 and 850~$\mu$m together with
1.2-mm data from the Institute de Radio Astronomie Millimetrique (IRAM)
30-m telescope.
They found the temperature to be in the range of 6 -- 16~K. 
All of these independent estimates are consistent with our findings.

\subsection{Column density}\label{column_density}

The measured values of the 850~$\mu$m flux density at the centre 
of the filament and in its outer part
are 0.082 Jy/beam and 0.022 Jy/beam 
respectively. The flux density of the outer part 
was measured 100 arcsec 
from the centre of the filament. From these measurements, together with 
the temperatures determined above, we can calculate the column density of 
material along the two lines of sight.

The mass per beam can be calculated using:
\begin{equation}
M_{beam}=\frac{S_{850}D^2}{\kappa_{850}B_{850,T}},
\end{equation}
where $S_{850}$ is the 850~$\mu$m flux density, $D$ is the distance to 
the filament, $\kappa_{850}$ is the mass opacity of the gas and dust, 
and $B_{850,T}$ is the value of the blackbody
function at a wavelength of 850~$\mu$m at temperature $T$. We assume a 
distance to TMC1 of 140~pc \citep{1978ApJ...224..857E}, and a mass 
opacity of ${\rm 0.01~cm^2g^{-1}}$ \citep[see][for a detailed discussion 
of both this value of $\kappa_{850}$ in particular and of this method of 
obtaining masses in general]
{1993ApJ...406..122A,1996A&A...314..625A,1999MNRAS.305..143W}.

The column density, $N(H_2)$, is then calculated using:
\begin{equation}
N(H_2) = \frac{M_{beam}}{2.8m_H \pi r_{beam}^2 },
\end{equation}
where $r_{beam}$ is the radius of the beam at the distance of TMC1. 
This yields column densities of $3.8 \times 10^{22}~{\rm cm}^{-2}$ 
and $4.4 \times 10^{21}~{\rm cm}^{-2}$ for the central part and the outer
part of the filament respectively. Hence we see that the centre of the 
filament has roughly an order of magnitude higher column density than
the outer part.

We can also calculate the column density using the 70-$\mu$m data, where 
the filament is seen in absorption, using the method described by Bacmann 
et al. (2000 -- see also Ward-Thompson et al. 2006). This compares the 
measured intensity where the absorption is the strongest ($I_{on}$), with 
the intensity off the filament ($I_{off}$). A third measure ($I_{fore}$) 
away from the region is also required to account for the zodiacal light and 
other foreground emission between us and the molecular cloud, as well as the 
large-scale background Galactic emission. The optical depth ($\tau_\lambda$) 
of the filament is given by the following \citep{2006MNRAS.369.1201W}:
\begin{equation}
e^{-\tau_{\lambda}} = \frac{I_{on} - I_{fore}}{I_{off} - I_{fore}} .
\end{equation}

The column density is calculated using:
\begin{equation}
N(H_2) = \frac{\tau_{\lambda}}{\sigma_{\lambda}},
\end{equation}
where $\sigma_\lambda$ is taken to be $2.6 \times 10^{-24}~\rm{cm^2}$ 
at 70~$\mu$m \citep[][see their figure~9]{1984ApJ...285...89D}.
Measured values of $I_{on}$, $I_{off}$ and $I_{fore}$ are 50.91, 51.13 
and 49.51 MJysr$^{-1}$ respectively, which lead to an
estimate of the column density of 
$5.6 \times 10^{22}~\rm{cm^{-2}}$. This is very close to the value 
of $3.8 \times 10^{22}~\rm{cm^{-2}}$
calculated above from the 850-$\mu$m data in emission.
The fact that two different data-sets from two different telescopes,
and two different methods of calculation, give essentially the same answer
give added confidence to our estimates.

\section{Radiative Transfer Modelling}\label{model}

In this section we employ more sophisticated techniques to model all of
the data, in order to understand the density structure of the filament
in more detail. We use a radiative 
transfer code \citep{2003A&A...407..941S} to model the data, using a variety
of parameter values to obtain the best fit simultaneously at all wavelengths.
In each case, the filament is approximated by a 
cylinder, which is heated externally
by the interstellar radiation field (ISRF). To determine 
the properties of the filament a large number of radiative transfer 
simulations were performed, and a best-fit solution was found.

We assume that the dust consists of carbonaceous and silicate grains with
size distribution from \citet{2001ApJ...548..296W}. These properties are
appropriate for the dust in the ISM \citep[$R_V=3.1$; see][]{2003ARA&A..41..241D}.

\subsection{Monte Carlo radiative transfer}

The radiative transfer calculations were performed using {\sc PHAETHON}, 
a 3D Monte Carlo radiative transfer code developed by  
\citet{2003A&A...407..941S}. The code uses a large number of monochromatic 
luminosity packets ($L$-packets)
to represent the radiation sources in the system. The $L$-packets
are injected into the cloud and interact (are absorbed, 
re-emitted, scattered) stochastically with it.  

If an $L$-packet is absorbed 
its energy is added to the local region and raises the local temperature. To 
ensure radiative equilibrium  the $L$-packet is re-emitted immediately with 
a new frequency chosen from the difference between the local cell emissivity 
before and after the absorption of the packet 
\citep{2001ApJ...554..615B,2005NewA...10..523B}.
For more details, see \citet{2003A&A...407..941S}.

We modelled the filament using a symmetric cylinder, and hence the code 
used here is optimised for the study of systems with axial symmetry. 
The filament is typically divided into $\sim4000$ cells. The number of 
cells used is chosen such that the density and temperature differences 
between adjacent cells are small.

\subsection{Heating of the filament}

Our simple modelling above has shown us that the centre of the filament is
colder than the outer parts. Thus we deduce that there are no heating
sources embedded within the filament, and hence the filament is starless.
In order to verify this, Spitzer MIPS 24 $\mu$m and Spitzer IRAC data for the filament 
were searched for YSOs and protostars \citep[c.f.][]{2004ApJS..154..363A, 2007ApJ...663.1149H,2007ApJS..171..447R}. With the exception of a single class I source
coincident with the outskirts of the filament, there are no observed internal heating sources. 
The only heating source considered is therefore the 
interstellar radiation field (ISRF). For this radiation field  we adopt a 
revised version of the \citet{1994ASPC...58..355B} interstellar radiation 
field (BISRF). The BISRF consists of radiation from giant stars and dwarfs, 
thermal emission from dust grains, cosmic background radiation, and 
mid-infrared emission from transiently heated small PAH grains 
\citep{2003cdsf.conf..127A}. 

This radiation field is modulated by the ambient cloud around
the filament. Hence the incident radiation field on the 
filament is enhanced at FIR and longer wavelengths, and attenuated at 
shorter wavelengths. The $L$-packets representing the ambient radiation 
field (typically a few $10^{10}$ packets) are injected from the outside 
of the virtual ambient cloud surrounding the filament with injection 
points and injection directions chosen to mimic an isotropic radiation field.

\subsection{Plummer-like geometry}\label{plummer_geometry}

\begin{figure}
\includegraphics[angle=270,width=87mm]{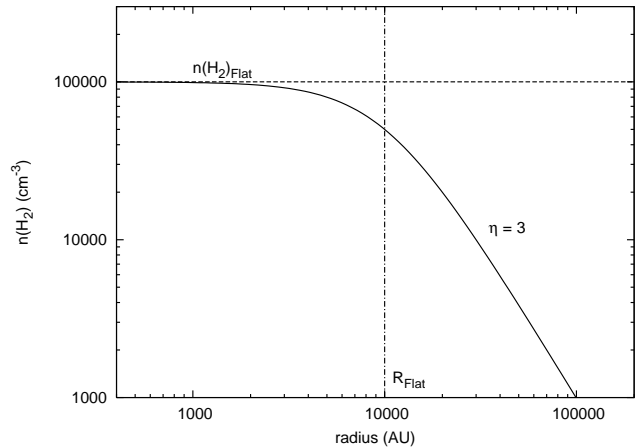}
\caption{A Plummer-like analytic density profile, with 
$n(H_2)_{flat}=10^{5}$~cm$^{-3}$, $R_{flat}~=~10^{4}$~AU and $\eta~=~3$.}
\label{plummer_diag}
\end{figure}

The geometry we assume is a cylinder with a 
Plummer-like density profile \citep{1911MNRAS..71..460P}. So 
we use axial symmetry with the density being approximately flat 
in the centre of the filament and dropping as a power-law in the 
envelope of the filament, according to the Plummer relation:
\begin{equation}
n(H_2)(r)=n(H_2)_{flat} \left[ \frac{R_{flat}}{(R_{flat}^2 + r^2)^{1/2} } 
\right]^\eta,
\end{equation}

\noindent
where $r$ is the distance from the axis of symmetry of the cylinder, 
$n(H_2)_{flat}$ is the density at the axis of symmetry, $R_{flat}$ is the 
extent of the region in which the density is approximately uniform, and 
$\eta$ is the power-law slope at large values of $r$. This analytic form is 
illustrated in Figure~\ref{plummer_diag}.  The filament is surrounded by a 
virtual ambient cloud, which has a uniform density $n(H_2)_{amb}$. The role 
of this cloud is to modify the ambient radiation field that heats the 
filament externally.

We use the term {\it Plummer-like} because a true Plummer profile has 
$\eta=5$, whereas we treat this as a free parameter. A Plummer-like 
profile was selected because it provides a high-density inner region, 
which decreases to large radii, with a small number of parameters. 
The model can produce images of the source at any requested wavelength,
convolved to any resolution. From these images we can produce one-dimensional 
cuts perpendicular to the cylinder for comparison with the data.

\begin{figure*}
\includegraphics[angle=270,width=175mm]{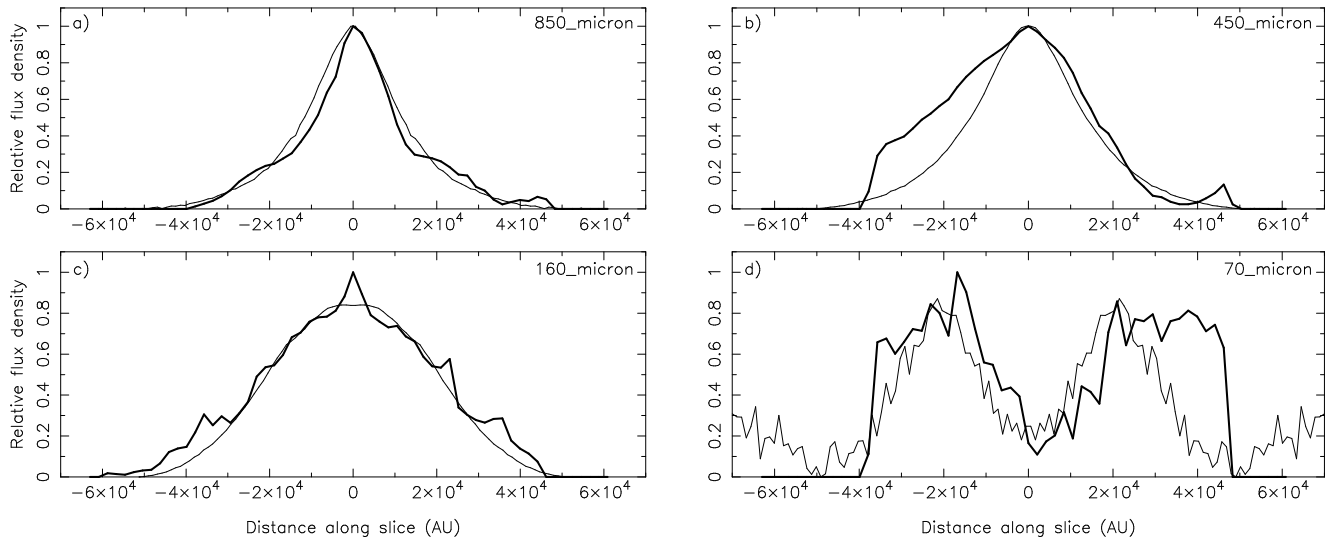}
\caption{The cuts through the data from Figure~\ref{slices} 
(bold lines), with overlaid (thin lines) the best fit profiles from
the radiative transfer model.}
\label{plummer_slices}
\end{figure*}

\citet{2001ApJ...547..317W} used a Plummer-like profile to model 
prestellar cores as they collapse to form protostars. They
found that this model successfully predicted 
both the observed infall rates, and the relative lifetimes of the 
different evolutionary stages.

The following parameter space was explored. $R_{flat}=5 \time 1 \times 10^2 - 1.5 \times 10^4$ AU, 
$n(H_2)_{flat}=7 \times 10^4 - 2 \times 10^7$ cm$^{-3}$, $\eta = 1 - 4$. 
In addition, the filament was embedded in a diffuse medium, with a visual extinction  
varying from $A_V = 0.01 - 0.55$.

Figure~\ref{plummer_slices} shows the best-fit results for this model.
The cuts through the model filament are compared to those 
through the data (see Figure~\ref{slices}). The parameters for this model 
are $R_{flat}=10^{4}$ AU, $n(H_2)_{flat}=1.75 \times 10^5$ cm$^{-3}$, $\eta = 3$ 
and $A_V$ in the external cloud equal to 0.39. The IRAS data are not included in this figure as
they are of insufficient quality for detailed modelling at this scale.

This model shows a good fit to the filament in emission at 850 and 160~$\mu$m, 
and fits one side of the asymmetric profile of the filament at 450~$\mu$m.
The model also shows the filament in absorption at 70~$\mu$m. 
While this is not necessarily a unique fit to the data, it shows that 
the filament can be modelled with a fairly simple geometry and uniform grain 
properties throughout. 

This is in contrast to the result of \citet{2003A&A...398..551S},
who need to invoke a non-uniform grain distribution in order to fit the far-infrared 
density profiles of a similar filament in Taurus. 
\citeauthor{2003A&A...398..551S} conclude that grain aggregation is important in
the central $3 \times 10^4$~AU, and is required to model the emission. We can explain the 
emission from our filament without a need for a change in grain properties.

\begin{figure}
\includegraphics[angle=270,width=87mm]{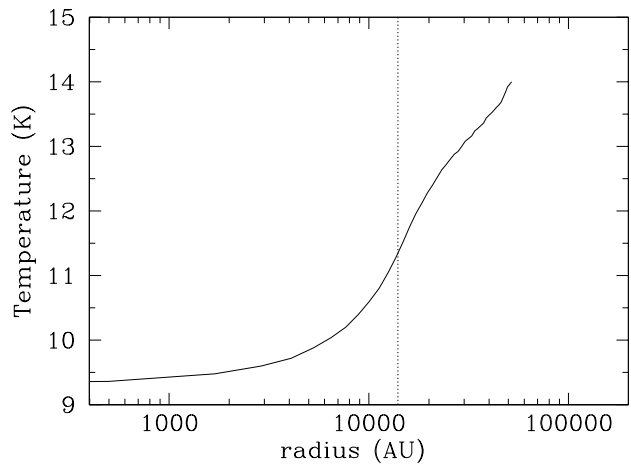}
\caption{The temperature profile from the best-fit model. The vertical dotted line shows the radial distance of the outer component of the filament (see section \ref{temp_analysis}.)}
\label{temperature_profile}
\end{figure}

The radiative transfer model allows us to measure the temperature of the dust grains as a function of position. This temperature does not vary with grain size, but is the same for all grains at a specific position. The temperature profile of the best-fit model is shown in Figure \ref{temperature_profile}. This gives us an independent temperature estimate for comparison with the SED fitting carried out in Section \ref{temp_analysis}. The model predicts a temperature of 12~K for the outer component of the filament, in agreement with the temperature determined from the SED fits. The central temperature is slightly higher than the best fit SEDs (9.8~K compared to $8 - 9$~K).  

\section{Conclusions}

We have presented submillimetre data from SCUBA on the JCMT and far-infrared 
data taken by Spitzer and IRAS, for a filament in TMC1 which we have
named the bull's tail. TMC1 is the densest part 
of the Taurus molecular cloud region.
The filament is seen in emission at all 
wavelengths longer than 100~$\mu$m, and in absorption at
all shorter wavelengths. 

The filament is significantly narrower at 850~$\mu$m, which we attribute 
to a narrow, dense inner filament, running through the centre of the broader 
and warmer main filament. 
We have shown intensity cuts through the filament
at wavelengths of 850, 450, 160, 100, 70, 
60, 25 and 12~$\mu$m, which are consistent with this scenario.

We have fitted SEDs to both the inner and outer components of the filament and 
find that the temperature decreases from 12~K in the outer part to 8~K at 
the centre. We then used a radiative transfer code to fit the profile of the filament. 
We modelled the filament using a Plummer-like density profile with constant grain 
parameters throughout, and find good fits to all wavelengths. 

\section*{Acknowledgements}
The authors would like to thank S. Schnee for helpful comments on this manuscript. This research is based on observations obtained with the James Clerk Maxwell Telescope, which is operated by the Joint Astronomy Centre in Hilo, Hawaii on behalf of the parent organizations STFC in the United Kingdom, the National Research Council of Canada and The Netherlands Organization for Scientific Research. The data were obtained using the Canadian Astronomy Data Centre, which is operated by the Herzberg Institute of Astrophysics, National Research Council of Canada. The observations were taken under the programs M98BI19, M98BC14, M99BC30, M01AI05, M01BC30, M01BN13 and M04BN16. This work is also based on observations made with the Spitzer Space Telescope, which is operated by the Jet Propulsion Laboratory, California Institute of Technology under a contract with NASA.  We acknowledge the use of NASA's {\em SkyView} facility (http://skyview.gsfc.nasa.gov) located at NASA Goddard Space Flight Center. DN, JMK and DS acknowledge STFC for PDRA support.

\end{document}